# Direct observation of displaced phonons responsible for small polaron charge transport in $(La_{1/3}Sm_{2/3})_{0.67}Ba_{0.33-x}Sr_xMnO_3$ (x = 0.0, 0.1, 0.2 and 0.33)

Saket Asthana[1], D. Bahadur[1] and C. M. Srivastava[2]

[1.] Department of Metallurgical Engineering and Materials Science, Indian Institute of Technology, Mumbai 400076, India

[2.] Department of Physics, Indian Institute of Technology, Mumbai 400076, India

**Abstract**

It is shown that charge transport in substituted manganites $(La_{1/3}Sm_{2/3})_{0.67}Ba_{0.33-x}Sr_xMnO_3$ (x = 0.0, 0.1, 0.2 and 0.33) is through polarons which are formed through electron-phonon coupling whose strength determines the mobility of the charge carrier. For weak coupling the free phonon mode is only slightly affected but in strong coupling the free phonon mode disappears and 'displaced' phonon modes appear. This is seen directly in IR-spectra.

The nature of the charge carriers and the mechanism of transport in $(R_{1-x}M_x)MnO_3$ manganites (R = Nd, Pr, Sm, La and M = Ba, Sr, Ca) has been extensively investigated [1-6] but there is still no theory that can satisfactorily account for the complex correlation between the magnetic and transport properties of the system. Since most attempts to understand the physics of the problem relate to the competition between the double exchange and superexchange [7-11], the fraction x depending on the concentration of the divalent alkaline earth metal atom becomes important as it governs the dominance of one interaction over the other. For example, the half doped manganites (x = 0.5) which are at the threshold of ferromagnetic to antiferromagnetic phase transition at low temperature comprise of regions with three different microscopic magnetic phases and two crystallographic structures simultaneously coexisting [12]. To explore the nature of the charge carriers we have therefore chosen x = 0.33. In this case, there is a single magnetic phase transition at $T_c$ as it is cooled from 300 K to low temperature and there is no first order phase transition of crystallographic phase. However, there is a strong dependence of the mobility of the charge carriers on temperature depending on the average ionic radius of the A site ion, $<r_A>$. Amongst others this is also reported by

Garcia-Munoz et al [13] who have observed in $(R_{1-x}R'_x)_{2/3}A_{1/3}MnO_3$ (M = Ca, Sr) variation of residual resistivity, ρ(0) by six orders ($10^{-3}$ to $10^3$ Ω cm) when average A-site ion radius, $<r_A>$, decreases from 1.2 Å to 1.11 Å.

Ramakrishnan [14] has attempted to account for it by empirically adding a factor $\exp[\lambda\delta M/M_s]$ to Mott's minimum resistivity, $\hbar a/e^2$, where a is the lattice constant and the expression is based on the uncertainty relationship for hopping charge carriers, giving $\rho_{min} \sim 10^{-3}$ Ω cm observed in manganites. Here δM is the deviation from the saturation magnetization, $M_s$, and λ is a large constant. Though the exponential term is based on experimental observation, the magnetic origin for localization is difficult to understand. We show that a similar result is obtained in $(La_{1/3}Sm_{2/3})_{0.67}Ba_{0.33-x}Sr_xMnO_3$ as x is varied from 0 to 0.33. The residual resistivity varies from 0.6 Ω-cm to $10^6$ Ω-cm with $<r_A>$ increasing from 1.15 Å to 1.20 Å. We show below that the variation in the magnitude of the residual resistivity, ρ(0), with $<r_A>$ is mainly due to the polaron-polaron interaction energy [15]. In this model the active free phonon that is coupled to the electron forming the small polaron has its frequency shifted to that of displaced phonons. In the weak electron-phonon coupling, the change in free phonon spectrum is small but in the case of strong coupling it is clearly noticeable in IR spectra. We further show that the frequency of the active phonon mode obtained from the resistivity data agrees with those obtained from IR measurement for the displaced phonons. Probably it is the first report of the observation of the displaced phonons along with those of free phonons from which these are formed.

**Theory**

We use the analysis of the polaron and displaced phonons based on the Holstein Hamiltonian used in ref. (16)

$$H = \sum_i \varepsilon_i c_i^+ c_i + \sum_{i,j} t_{ij}(c_{i+j,\sigma}^+ c_{i,\sigma} + HC) + \sum_q \omega_q a_q^+ a_q + \sum_{q,i} g_{iq} c_i^+ c_i (a_q^+ + a_q) \quad (1)$$

Here $t_{ij}$ is the electron transfer integral, $g_{iq}$ is the electron-phonon coupling constant, $\varepsilon_i$ is electron energy at site $i$, $\omega_q$ is the energy of free phonon, and $c_i(c_i^+)$ and $a_i(a_i^+)$ are electron and phonon annihilation (creation) operators respectively. The polaron canonical transformation yields [15, 16]

$$H_{eff} = \sum_i (\varepsilon_i - \varepsilon_p) l_i^+ l_i + \sum_{i,j} (t_{ij} l_i^+ l_j X_{ij} + HC) + \sum_q \omega_q b_q^+ b_q \qquad (2)$$

$$\varepsilon_p = \sum_q \langle g_{iq}^2 \rangle / \omega_q \qquad (3)$$

where $l_i^+$ creates a polaron at site $i$, $X_{ij}$ are the Frank-Condon transition between sites $i$ and $j$, $\varepsilon_p$ is the polaron stabilization energy and $b_q^+$ is a creation operator for displaced phonons, the displacement determined by the position of all electrons

$$b_q^+ = a_q^+ + \sum_i g_{iq} e^{iq \cdot R_i} c_i^+ c_i \qquad (4)$$

The detailed analysis of Frank-Condon transitions shows that polarons at one site interacts with that at the neighbouring site and hopping can occur only if one site is occupied and the other is vacant. This gives a factor $\text{sech}^2(\varepsilon_p/2T)$ in the mobility of the polaron. The other temperature dependent factor that enters the mobility is the activation energy $U_0 = (1/4) n_{ph} \omega_{ph}$ where $n_{ph}$ is the average number of phonons in the polaron cloud. For the electron to jump from site $R_i$ to $R_j$ this adds a factor $\exp(-U_0 \langle \sin^2 \frac{1}{2} \underline{q} \cdot \underline{a} \rangle) = \exp(-U_0 \xi^2/T)$ where $\underline{q}$ is the wave vector and $\underline{a} = \underline{R}_i - \underline{R}_j$ is the displacement vector for the jump [16,17]. Following Kubo's current-current correlation formulation Reik [16] has obtained $H_{eff}$ to obtain an expression for dc resistivity which has been adopted by one of us for transport in manganites using the correlated polaron approximation [17],

$$t_{ij} = t = \omega_{ph}, \; 1/\tau = \bar{\theta}_D = \omega_{ph} \langle \sin^2 \frac{1}{2} \underline{q} \cdot \underline{a} \rangle = \omega_{ph} \xi^2 \qquad (5)$$

$$\rho_c^{hop} = \frac{AT}{n} \left[ 1 + c(1 - m^2(t)) \sigma_a^2 \right] \cosh^2(\varepsilon_p / 2T) \exp[U_0 (1 - m^k) \sigma_a^2 / T], \; T > \bar{\theta}_D / 4 \qquad (6)$$

with $m(t) = M(T)/M(0)$, n the number density of charge carriers, and $c$ and $k$ are constants depending on the strength of the spin-spin scattering in polaron transport. Here $A = 2k_B / \sqrt{\pi} \omega_{ph} a^2 e^2$, $\omega_{ph}$ is the frequency of the active phonon, a is the lattice constant, $\sigma_a$ is the short-range atomic order parameter that varies with T as $(1 - 0.75 t_{ca}^3)^{1/2}$ where $t_{ca} = T/T_{ca}$. Equation (6) is similar to that obtained for glassy semiconductor ($m(t) = 0$) by Mott [18] for $T > \bar{\theta}_D / 4$. For $T < \bar{\theta}_D / 4$, quantum effects dominate over classical effects and according to Mott, due to processes like variable range hopping and Anderson localization, as T tends to zero, ρ increases as $A'\exp(b/T^{1/4})$. This is clearly not applicable to manganites where for the same number of charge carriers (x = constant) and nearly identical room temperature resistivity, the residual resistivity, ρ(0), changes from 2 mΩ-cm to $10^3$ Ω-cm, six orders of magnitude, when $<r_A>$ is varied [13,14]. We show that the residual resistivity is proportional to $\cosh^2(2\varepsilon_p/\bar{\theta}_D)$ where $\varepsilon_p$ is the polaron energy in eq. (6) and $\bar{\theta}_D$ is the Debye temperature relating to the displaced phonon given by eq. (5)

Since scattering by phonons is proportional to lattice strain $\langle\delta R^2\rangle/R^2$ which is equal to $1.6 k_B T/Ms^2$ in the high temperature limit and to $0.4 k_B \theta_D/Ms^2$ in the low temperature limit where M is the mass of the ion and s is the velocity of sound [19], we assume, following Mott, that eq. (6) gives the resistivity for $T < \bar{\theta}_D/4$, by replacing T by $\bar{\theta}_D/4$ in the prefactor and the cosh term,

$$\rho_c^{hop} = \frac{A\bar{\theta}_D}{4n}\left[1+c\left(1-m^2(t)\right)\sigma_a^2\right]\cosh^2(2\varepsilon_p/\bar{\theta}_D)\exp[U_0(1-m^k)\sigma_a^2/T], \quad T < \bar{\theta}_D/4 \qquad (7)$$

For T→0 using c = 1 and spin-wave theory,

$$\rho_c^{hop} = \rho(0)\left[1+2\zeta T^{3/2}\sigma_a^2\right] \qquad (8)$$

where $\zeta$ is the spin-wave constant and

$$\rho(0) = \frac{A\bar{\theta}_D}{4n}\cosh^2\left(\frac{2\varepsilon_p}{\bar{\theta}_D}\right) \qquad (9)$$

With the same parameters A/n, $T_c$, $T_{ca}$, $\bar{\theta}_D$, $U_0$ and k = 2.3 we have plotted two curves for ρ vs T, one using eq. (6) that applies for $T > \bar{\theta}_D/4$ and the other using eq. (7) for $T < \bar{\theta}_D/4$ that fit the observed ρ(T) curves best. This is shown in Fig. 1 and the parameters are given in Table 1. First the value of the $\varepsilon_p$ was estimated from the fitted plot in $T > \bar{\theta}_D/4$ region. Values of $\bar{\theta}_D$ are then estimated using eq. (9) from the values of $\varepsilon_p$ and ρ(0). In Fig. 2 is shown the variation of ρ(0) with $<r_A>$ for our data and data of ref. [13] for $(R_{1-x}R'_x)_{2/3}A_{1/3}MnO_3$ (M = Ca, Sr). The linear nature of the curves is explained on the basis of eq. (9) if $4\varepsilon_p/\bar{\theta}_D$ α ln ρ(0), is linearly dependent on $<r_A>-<r_A>_c = \Delta$, where $<r_A>_c$ is the value for the structure closet to cubic perovskite. As $\Delta$ increases, at T→0 the polarons change from itinerant to localized states since Mn-O-Mn angles deviate more from $180^0$. From Fig. 2, $\Delta$ varies linearly with $4\varepsilon_p/\bar{\theta}_D$. Fig. 3 gives the vibrational spectra of the samples. It is shown that a magnetic and charge-order Bravais lattice exists for $A^{3+}_{1-x}B^{2+}_x Mn^{3+}_{1-x} Mn^{4+}_x O_3$ for x = n/8 where n is an integer and the cell constant is 2a, double that of the chemical cell [17]. Correlated polarons transport in manganites with x = 1/3 occurs with the lattice vibrational mode with frequency $\omega_{ph} = 5.5 \times 10^{12}$ Hz [17]. This nearly corresponds to the 179 cm$^{-1}$ bond stretching mode in which the two oxygen atoms move along the bond direction ± z in the $MnO_6$ octahedra with all other four oxygen atoms surrounding the Mn atom remaining silent [Fig. 4a]. From Table 1, $\bar{\theta}_D/\omega^1_{ph} = 0.42$, 0.47 and 0.74 for x = 0, 0.1 and 0.2, respectively, so from eq.(5) and $\left\langle \sin^2 \frac{1}{2}\underline{q}.\underline{a} \right\rangle$ lies between 0.4 and 0.74. On the other hand, for x = 0.33 which has largest mobility as T→0 has a value of $\bar{\theta}_D/(\omega^1_{ph}+\omega^2_{ph})$ equal to 0.72. In this case both the modes of vibration, the MnO stretching ($\omega^1_{ph}$ ~ 179 cm$^{-1}$) and bending plus stretching ($\omega^2_{ph}$ ~ 395 cm$^{-1}$) are similar to that observed for $SrTiO_3$ [20] appear to be active (Fig. 4). In the Bravais lattice of cell constant 2a, $\omega^1_{ph}$ and $\omega^2_{ph}$ modes could exist amongst edge-sharing octants and $\bar{\theta}_D = 0.72(\omega^1_{ph}+\omega^2_{ph})$.

Fig. 3 shows the disappearance of the phonon modes $\omega^1_{ph}$ and $\omega^2_{ph}$ in the present system for x = 0.33 due to strong electron-phonon coupling $g_{iq}$ in eq. (4) which leads to the appearance of displaced phonons that play a dominant role in the charge transport process. From eq. (9) the mobility as T→0 is proportional to sech$^2(2\varepsilon_p/\bar{\theta}_D)$ and hence is exponentially dependent on $4\varepsilon_p/\bar{\theta}_D$ for $\varepsilon_p/\bar{\theta}_D \geq 1$. As $\varepsilon_p$ is nearly independent of x it is $\bar{\theta}_D$ that controls the low temperature transport, lower the $\bar{\theta}_D$, the lower is the mobility and hence higher the residual resistivity.

**Table.1**

| x | $T_{ca}$ (°K) | $A/n$ ($10^{-3}$/cc) | $\varepsilon_p$ (°K) | $\bar{\theta}_D$ (°K) | $U_0$ (°K) | $k$ | $n$ (x $10^{18}$/cc) |
|---|---|---|---|---|---|---|---|
| 0 | 310 | 1.8 | 500 | 115 | 420 | 2.3 | 4.1 |
| 0.1 | 310 | 1.8 | 470 | 125 | 500 | 2.3 | 4.1 |
| 0.2 | 310 | 1.0 | 450 | 200 | 600 | 2.3 | 7.4 |
| 0.33 | 310 | 0.6 | 380 | 430 | 750 | 2.3 | 12.4 |

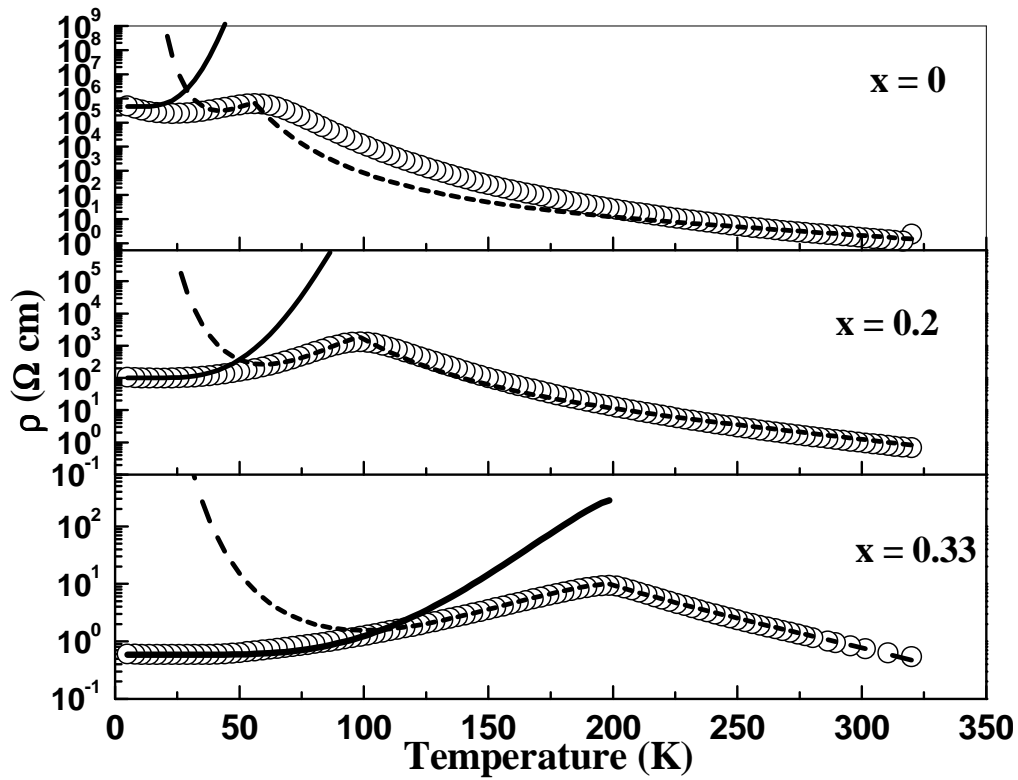

**Fig.1**

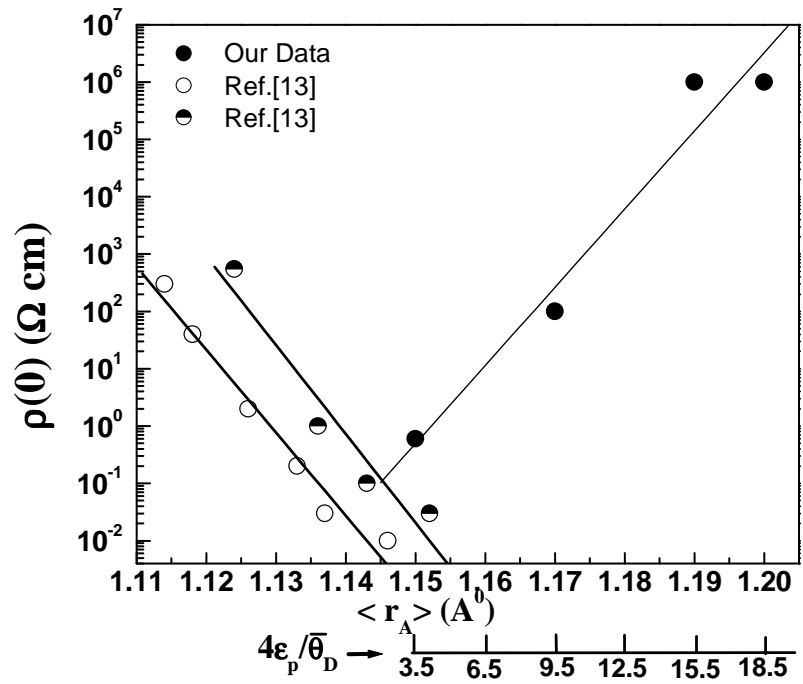

**Fig.2**

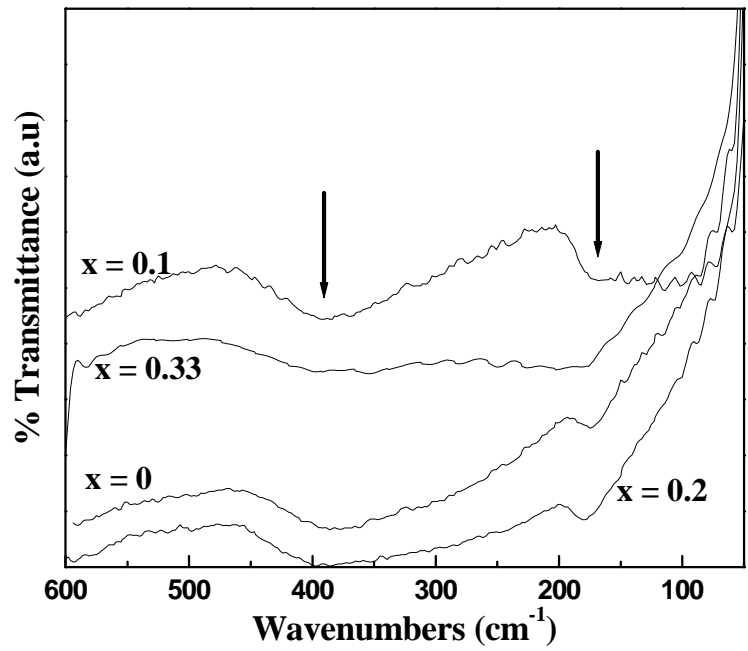

**Fig. 3**

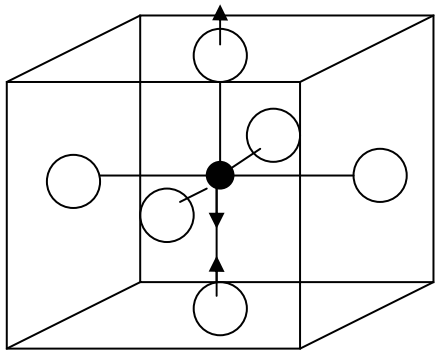 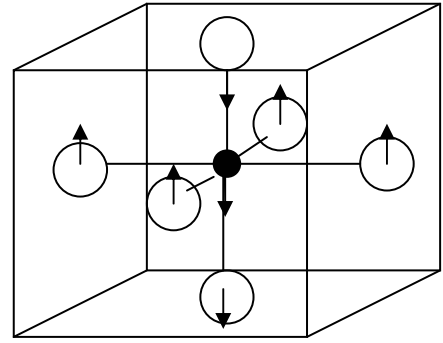

(a) $\omega^1_{ph}$ (179 cm$^{-1}$)    (b) $\omega^2_{ph}$ (395 cm$^{-1}$)

**Fig. 4**

**Figure Captions**

**Fig.1.** Observed $\rho(T)$ curves for $(La_{1/3}Sm_{2/3})_{2/3}Ba_{1/3-x}Sr_xMnO_3$ (x = 0, 0.2 and 0.33) are shown in hollow circles. The fitted plots are shown by dotted lines (- - -) for T > $\bar{\theta}_D/4$ eq. (5) and by solid lines (——) for $0 < T \leq \theta_D/4$ eq. (7).

**Fig.2.** Variation of $\rho(0)$ with $<r_A>$ for $(La_{1/3}Sm_{2/3})_{2/3}Ba_{1/3-x}Sr_xMnO_3$ (x = 0, 0.1, 0.2 and 0.33) and data for $(R_{1-x}R'_x)_{2/3}A_{1/3}MnO_3$ (M = Ca (hollow circles), Sr (half filled circles)) taken from Ref. [13]. The values of $4\varepsilon_p/\bar{\theta}_D$ for our data using eq. (9) and Table 1 are also given.

**Fig.3.** IR spectra of $(La_{1/3}Sm_{2/3})_{2/3}Ba_{1/3-x}Sr_xMnO_3$ (x = 0, 0.1, 0.2 and 0.33). The arrows denote the ($\omega^1_{ph} \sim$ 179 cm$^{-1}$) and bending ($\omega^2_{ph} \sim$ 395 cm$^{-1}$) phonon peaks.

**Fig.4.** The MnO stretching ($\omega^1_{ph} \sim$ 179 cm$^{-1}$) and bending plus stretching ($\omega^2_{ph} \sim$395 cm$^{-1}$) vibration modes similar to those observed in SrTiO$_3$ [20] are shown for the present system.

**Table Captions**

**Table1.** The fitted parameters, $\varepsilon_p$, $\bar{\theta}_D$, $U_0$ and $n$ are given for the series $(La_{1/3}Sm_{2/3})_{2/3}Ba_{1/3-x}Sr_xMnO_3$ (x = 0, 0.1, 0.2 and 0.33). The value of $A/n$ has been obtained from the $\rho(0)$ values using $\omega^1_{ph}$ with a = 3.858 Å.